\documentclass[aps,prd,twocolumn]{revtex4-1}

\usepackage{graphicx}  % needed for figures
\graphicspath{ {images/} } %the images are kept in a folder named images under the current directory.
\usepackage{latexsym}
\usepackage{slashed}
\usepackage{dcolumn}   % needed for some tables
\usepackage{bm}        % for math
\usepackage{amsmath}
\usepackage{amssymb}   % for math
\usepackage{longtable}
\usepackage{float}
\newcommand{\be}{\begin{equation}}
\newcommand{\ee}{\end{equation}}
\newcommand{\bea}{\begin{eqnarray}}
\newcommand{\eea}{\end{eqnarray}}
\newcommand{\bma}{\begin{matrix}}
\newcommand{\ema}{\end{matrix}}

\newcommand{\bml}{\begin{mathletters}}
\newcommand{\eml}{\end{mathletters}}
\newcommand{\bes}{\begin{subequations}}
\newcommand{\ees}{\end{subequations}}

\newcommand{\bi}{\begin{itemize}}
\newcommand{\ei}{\end{itemize}}
\newcommand{\gev}{~{\rm GeV}}

\newcommand{\mev}{~{\rm MeV}}

\begin{document}
\title{What does a non-vanishing neutrino mass have to say about the strong CP problem?}
\author{P. Q. Hung}
\email{pqh@virginia.edu}
\affiliation{Department of Physics, University of Virginia,
Charlottesville, VA 22904-4714, USA}

\date{\today}

\begin{abstract}
A new solution to the strong CP problem with distinct experimental signatures at the LHC is proposed. It is based on the Yukawa interactions between mirror quarks, Standard Model (SM) quarks and Higgs singlets. (Mirror quarks and leptons which include non-sterile right-handed neutrinos whose Majorana masses are proportional to the electroweak scale, form the basis of the EW-$\nu_R$ model.) The aforementioned Yukawa couplings can in general be complex and can contribute to $Arg\, Det M$ ($\bar{\theta} = \theta_{QCD} + Arg\, Det M$) at tree-level. The crux of matter in this manuscript is the fact that {\em no matter how large} the CP-violating phases in the Yukawa couplings might be, $Arg\, Det M$ can remain small i.e. $\bar{\theta} < 10^{-10}$ for reasonable values of the Yukawa couplings and, in fact, vanishes when the VEV of the Higgs singlet (responsible for the Dirac part of the neutrino mass in the seesaw mechanism) vanishes. The smallness of the contribution to $\bar{\theta}$ is {\em principally due} to the smallness of the ratio of the two mass scales in the seesaw mechanism: the Dirac and Majorana mass scales.
%Several prominent proposals have been put forward as solutions to the Strong CP problem, among which are the Axion model and models with new extra quarks.  

\end{abstract}

\maketitle

%\section{Introduction}

It is a well-known fact that, although CPT appears to be respected as a symmetry of nature, CP and T are not as weak interaction experiments have shown us. Furthermore, studies of the QCD vacuum, the so-called $\theta$-vacuum, revealed that an additional CP-violating term is added to the Lagrangian in the form $\theta_{QCD} \, (g_{3}^2/32 \pi^2) G_{a}^{\mu \nu} \tilde{G_{\mu \nu}^{a}}$. In addition, the electroweak sector contributes another similar term through quark mass matrices so that the total $\theta$ is now $\bar{\theta} = \theta_{QCD} + Arg Det M$.

Constraints coming from the absence of the neutron electric dipole moment give $\bar{\theta} < 10^{-10}$ \cite{theta}. This is the famous strong CP problem: why $\bar{\theta}$ which is the sum of the contributions from the strong and weak sectors should be so small. Several lines of approach toward a solution to the strong CP problem have been proposed. The most famous one is the Peccei-Quinn axion \cite{peccei} where a new global $U(1)_{PQ}$ was added and where $\bar{\theta}$ is driven dynamically to zero. The axion is still elusive and its search is going on. A very early class of models solving the strong CP problem without the axion and using either soft CP breaking or simply P and T invariance can be found in \cite{mohapat}. Another line of approach \cite{SBCP}  is to assume CP conservation of the Lagrangian so that $\bar{\theta}=0$ at tree level and to postulate the existence of heavy fermions (within a Grand Unification context such as SU(5) or an extended gauge group $SM \times G$) to generate a non-vanishing $\bar{\theta}$ at loop levels. In this class of models, CP is spontaneously broken giving rise to potential problems with issues such as domain walls. 
%In another class of models \cite{mohapat}, $\bar{\theta}=0$ at tree-level and one loop.

Our approach to the strong CP problem is a non-axionic one and is more similar in spirit to the approach which assumed the presence of non-SM fermions, except for a few crucial differences. It {\em does not impose} CP conservation of the Lagrangian.Three questions that need to be addressed are the following: 1) If CP conservation is not imposed on the Lagrangian, what symmetry allows us to set the QCD $\theta_{QCD}$ to be equal to zero at tree level?; 2) Since CP can explicitly be violated by the complex Yukawa couplings, what prevents $Arg Det M$ from exceeding the upper bound of $10^{-10}$?; 3) Last but not least, can the solution be found {\em solely} within the gauge structure  of the SM, namely $SU(3) \times SU(2) \times U(1)$? 

The answers to the aforementioned three questions can be found in the EW-$\nu_R$ model \cite{nur} and will be elaborated below. All the ingredients for a solution to the strong CP problem are already contained in this model. However, a few key points about the EW-$\nu_R$ model need to be mentioned: 1) it avoids the Nielsen-Ninomiya no-go theorem \cite{nielsen} (which says that one cannot put the SM on the lattice without having mirror fermions interacting with the same W and Z bosons) by postulating the very existence of these mirror fermions. Note that mirror fermions are also motivated within the framework of $E_6$ with $27_L$ and $27^{c}_L$ representations \cite{nur}; 2) the right-handed neutrinos, being part of a right-handed mirror lepton doublet, are now {\em non-sterile} (or {\em fertile}) and obtain Majorana masses which are proportional to the electroweak scale and can be produced at the LHC and searched for by looking for like-sign dileptons events; 3) it satisfies the electroweak precision constraints as well as being able to accommodate the 125-GeV scalar in an interesting way.

%But first, a little exercise on a simplified model is in order for the purpose of illustrating the assertion made above. 
We start out with the one-generation case in the EW-$\nu_R$ model. 
%where there is no CKM but where CP violation is explicit in the Yukawa couplings between SM and mirror quarks. 
This helps to separate the two issues, that of the strong CP violation and that of the weak CP violation present in the CKM matrix for three (or more) generations of quarks.

 The gauge group is $SU(3)_C \times SU(2)_W \times U(1)_Y$. We have one generation of SM quarks: $q_L= (u_L,d_L), u_R, d_R$, and one generation of mirror quarks: $q_R^M= (u_R^M, d_R^M), u_L^M, d_L^M$. (The full model can be found in the EW-$\nu_R$ model \cite{nur}.) For the purpose of this manuscript, we will focus on the Yukawa couplings:
 \bea
 \label{mass}
 {\cal L}_{mass}& = & g_u \bar{q}_L \tilde{\Phi}_2 u_R + g_d \bar{q}_L \Phi_2 d_R + g_{u^M} \bar{q^M}_R \tilde{\Phi}_{2M} u^M_L \nonumber \\
 &&+ g_{d^M} \bar{q^M}_R \Phi_{2M} d^M_L +H.c. \,,\,
 \eea
 where $\tilde{\Phi}_{2,2M}  \equiv \imath \tau_2 \Phi_{2,2M}$ with $\Phi_{2,2M}$ being the two Higgs doublets of the extended EW-$\nu_R$ model \cite{125}.
 \bea
 \label{mixing}
 {\cal L}_{mixing} &=& g_{Sq}\bar{q}_L \phi_S \bar{q^M}_R + g_{Su}\bar{u}^M_L \phi_S  u_R  \nonumber \\
 &&+ g_{Sd}\bar{d}^M_L \phi_S  d_R + H.c. \,,
 \eea
 where $\phi_S$ is a Higgs singlet. The rationale for introducing the aforementioned degrees of freedom can be found in \cite{nur}.
 
% Without loss of generality, we will assume that $g_u$, $g_d$, $g_{u^M}$ and $g_{d^M}$ are real in order to give real masses and that $g_{Sq}$, $g_{Su}$ and $g_{Sd}$ are complex so that CP can be violated explicitly in Eq.~\ref{mixing}. 
Notice that $g_u$, $g_d$, $g_{u^M}$, $g_{d^M}$, $g_{Sq}$, $g_{Su}$ and $g_{Sd}$ can, in general be complex. If we absorb the phases into $u_R$, $u^{M}_L$, $d_R$ and $d^{M}_L$ to make the {\em diagonal} elements of the ($2 \times 2$) up and down mass matrices {\em real} then the {\em off-diagonal} elements stay {\em complex}.
 Furthermore, a global symmetry was invoked in \cite{nur,125} to ensure that the Yukawa couplings take the form as shown in Eq.~(\ref{mass}). $\langle \Phi_2 \rangle = v_2$ and $\langle \Phi_{2M} \rangle = v_{2M}$ give non-vanishing masses to the SM and mirror quarks, namely $m_u$, $m_d$, $M_u$ and $M_d$ respectively. (From \cite{nur,125}, $v_2^2 + v_{2M}^2 + 8 v_M^2 = (246 \gev)^2$ where $v_M$ is the VEV of the Higgs triplet which gives the Majorana mass to $\nu_R$.)  The mass mixing between SM and mirror quarks comes from Eq.~(\ref{mixing}). Writing $g_{Sq} = |g_{Sq} | \exp(\imath \theta_q)$, $g_{Su} = |g_{Su} | \exp(\imath \theta_u)$ and $g_{Sd} = |g_{Sd} | \exp(\imath \theta_d)$ and with $\langle \phi_S \rangle = v_S$, we obtain the following mass matrices ($m_u$, $M_u$, $m_d$ and $M_d$ are real)
 
 \begin{equation}
	\label{Mu}
	{\cal M}_u=\left( \begin{array}{cc}
	m_u & |g_{Sq} | v_S \exp(\imath \theta_q) \\
	 |g_{Su} | v_S  \exp(\imath \theta_u)& M_u
	\end{array} \right) \,,
\end{equation}
 \begin{equation}
	\label{Md}
	{\cal M}_d=\left( \begin{array}{cc}
	m_d& |g_{Sq} | v_S \exp(\imath \theta_q) \\
	 |g_{Sd} | v_S  \exp(\imath \theta_d)& M_d
	\end{array} \right) \,.
\end{equation}

One can now compute $\bar{\theta}$, namely
\be
\label{thetabar}
\bar{\theta} = \theta_{QCD} + Arg Det ({\cal M}_u {\cal M}_d) \,.
\ee
Two important questions are in order.
\bi
\item Can $\theta_{QCD} $ be {\em zero} at tree level?

To answer this question, we make use of the distinct feature of the EW-$\nu_R$ model which is {\em parity invariance} above the electroweak scale coming from the fact that one has both left- and right-handed fermions transforming in the same way under $SU(2)_W$ (hence the subscript "W" instead of "L"). Since $G_{a}^{\mu \nu} \tilde{G_{\mu \nu}^{a}}$ is {\em odd} under Parity, it follows that the QCD $\theta_{QCD} =0$ at tree level. (A similar argument was used in the Left-Right symmetric model \cite{mohapat}.)

\item Since CP is explicitly violated in ${\cal M}_{u,d}$, could their {\em tree-level} contribution to $Arg Det ({\cal M}_u {\cal M}_d)$ be {\em naturally} small without fine-tuning the CP phases?

\ei

From Eq.~(\ref{Mu},\ref{Md}), we obtain ($C_u \equiv m_u M_u$, $C_d \equiv m_d M_d$, $C_{Su} \equiv  |g_{Sq}||g_{Su}| v_S^2$ and $C_{Sd} \equiv  |g_{Sq}||g_{Sd}| v_S^2$)
\bea
Arg Det ({\cal M}_u {\cal M}_d)&=&Arg \{ (C_u - C_{Su} \exp[\imath(\theta_q + \theta_u)]) \nonumber \\
&&(C_d - C_{Sd} \exp[\imath(\theta_q + \theta_d)])\} \,.
\eea
Neglecting the term proportional to $C_{Su} C_{Sd}$ since (as we shall explain below) $C_{Su} C_{Sd} \ll C_u C_d, C_{Su}C_d, C_{Sd} C_u$, we obtain with $\theta_{Weak} \equiv Arg Det ({\cal M}_u {\cal M}_d)$
\be
\label{thetaw1}
\theta_{Weak} \approx \tan^{-1} \frac{-(C_{Su} C_{d}\sin(\theta_q + \theta_u) + C_{Sd} C_{u}\sin(\theta_q + \theta_u))}{C_d C_u -C_{Su} C_{d}\cos(\theta_q + \theta_u)- C_{Sd} C_{u}\cos(\theta_q + \theta_u)}
\ee
Defining
\be
\label{ru}
r_u = \frac{C_{Su}}{C_u}=\frac{|g_{Sq}||g_{Su}| v_S^2}{m_u M_u} \,,
\ee
\be
\label{rd}
r_d= \frac{C_{Sd}}{C_d}=\frac{|g_{Sq}||g_{Sd}| v_S^2}{m_d M_d} \,,
\ee
Eq.~(\ref{thetaw1}) can now be put in a neater form
\be
\label{thetaw2}
\theta_{Weak} \approx \frac{-(r_u \sin(\theta_q + \theta_u) + r_d \sin(\theta_q + \theta_d))}{1- r_u \cos(\theta_q + \theta_u)-r_d \sin(\theta_q + \theta_d)}
\ee 

%We are now ready to discuss the implications of Eq.~(\ref{thetaw2}).
\bi
\item First, we notice from Eq.~(\ref{thetaw2}) that $\theta_{Weak} =0$ when the VEV of the singlet Higgs vanishes i.e. when $v_S=0$. This is valid for {\em any} value of the phases $\theta_{q,u,d}$.

\item $\theta_{Weak}$ can also vanish if all the phase angles vanish or if $\theta_q =-\theta_u=-\theta_d$. Since theses are special cases, we will not consider them here but will instead keep them arbitrary.

\item As shown in \cite{nur}, a non-vanishing value for $v_S$ implies a non-vanishing Dirac mass of the neutrino participating in the seesaw mechanism i.e. $m_{\nu} = m_D^2/M_R$. From \cite{nur}, $m_D = g_{Sl} v_S$ coming from the interaction $g_{Sl} \bar{l}_L \phi_S \l^M_R + H.c.$ where $l_L=(\nu_L, e_L)$ and $l^M_R =(\nu_R, e^M_R)$. Here $M_R$ is the Majorana mass of the right-handed neutrino coming from $g_M \, (l^{M,T}_R \ \sigma_2)(\imath  \tau_2) \ \tilde{\chi} \ l^M_R$ where $\chi$ is a triplet Higgs with $Y/2=1$ and whose VEV is $v_M$.

\item Since $M_R > M_Z/2 \sim 45 \gev$ (from the Z-width constraint), one gets $m_D < 100\, keV$ \cite{nur}. 

\item One can rewrite $r_u$ and $r_d$ as
\be
\label{newru}
r_u=(\frac{|g_{Sq}||g_{Su}|}{g_{Sl}^2})(\frac{m_D^2}{m_u M_u}) \,,
\ee
\be
\label{newrd}
r_d=(\frac{|g_{Sq}||g_{Sd}|}{g_{Sl}^2})(\frac{m_D^2}{m_d M_d}) \,.
\ee
 $r_{u,d} \ll 1$ if one assumes that $|g_{Sq}||g_{Su}| \leq g_{Sl}^2$,

\item It is interesting to notie that, since the Majorana mass of the right-handed neutrinos is also proportional to the electroweak scale $246 \gev$, $r_u$ and $r_d$ which will determine the size of $\theta_{Weak}$ have the following proportionality
$r_u \propto m_\nu / m_u; r_d \propto m_\nu / m_d$
and vanish as $m_\nu \rightarrow 0$.
\item One can now rewrite $\theta_{Weak}$ as
\be
\label{thetaw3} 
\theta_{Weak} \approx -(r_u \sin(\theta_q + \theta_u) + r_d \sin(\theta_q + \theta_d)) \,.
\ee

\item As discussed in \cite{125}, one expects the mirror quarks to be heavy. For the sake of estimation, we shall take $M_u \sim M_d \sim 400 \gev$. Furthermore, since we are dealing with the one-generation case, let us take the most extreme case, namely $m_u \sim 2.3 \mev$ and $m_d \sim 4 \mev$. With the constraint $m_D < 100\, keV$, one obtains the following bound
\bea
\label{bound}
\theta_{Weak} &<& -10^{-8}\{(\frac{|g_{Sq}||g_{Su}|}{g_{Sl}^2})\sin(\theta_q + \theta_u)  \nonumber \\
&&+ (\frac{|g_{Sq}||g_{Sd}|}{g_{Sl}^2})\sin(\theta_q + \theta_d)\}
\eea
\item What does the inequality (\ref{bound}) imply? $|\theta_{Weak}| < 10^{-10}$ regardless of the values of the CP phases.
Even if one had {\em maximal} CP violation in the sense that $\theta_q + \theta_u \sim \theta_q + \theta_d \sim \pi/2$, $|\theta_{Weak}| < 10^{-10}$ provided $|g_{Sq}| \sim |g_{Su}| \sim |g_{Sd}| \sim 0.1 g_{Sl}$. 

\item This has interesting phenomenological implications concerning the searches for mirror quarks and leptons at the LHC \cite{search}. In fact, constraints coming from $\mu \rightarrow e \gamma$ \cite{muegamma} and from $\mu$-$e$ conversion \cite{mu2e} indicate that $g_{Sl} < 10^{-4}$ which would imply in the present context that $|g_{Sq}| \sim |g_{Su}| \sim |g_{Sd}| < 10^{-5}$. This implies the possibility of observing the decays of mirror quarks and leptons from the process $f^M \rightarrow f + \phi_S$ (where $f^M$ and $f$ stand for mirror and SM fermions respectively) at {\em displaced vertices} (large decay lengths) because of the small Yukawa couplings.
 
\item As already pointed out in \cite{nur}, the mass mixing between SM and mirror quarks is tiny, being proportional to the ratio of neutrino to quark mass. For most practical purpose, the mass eigenstates are approximately pure SM and mirror states.
\ei

A full analysis will involve three generations and will be more complicated. As opposed to the one-generation case where we have a $2 \times 2$ matrix, we will now have a $6 \times 6$ matrix of the form
\begin{equation}
	\label{6x6u}
	{\cal M}_u=\left( \begin{array}{cc}
	M_u & M_{q_L q^M_R} \\
	 M_{u_R u^{M}_L}& M_{u^M}
	\end{array} \right) \,,
\end{equation}
\begin{equation}
	\label{6x6d}
	{\cal M}_d=\left( \begin{array}{cc}
	M_d & M_{q_L q^M_R} \\
	 M_{d_R d^{M}_L}& M_{d^M}
	\end{array} \right) \,,
\end{equation}
where each element of the above matrices are $3 \times 3$ matrices. The matrices $M_{q_L q^M_R}$, $M_{u_R u^{M}_L}$ and $M_{d_R d^{M}_L}$ contain matrix elements which are proportional to the VEV of the singlet Higgs field, namely $v_S$. As we have shown above for the one-generation case, these are much smaller than matrix elements of $M_u$, $M_d$, $M_{u^M}$ and $M_{d^M}$. For this reason, those mass matrices can be diagonalized separately, neglecting mixing. Furthermore, we believe that the result for $\theta_{Weak}$ will not be too different for that given in Eq.~(\ref{bound}).

We carried out an analysis based on a simplified version of the full model. (The full analysis is beyond the scope of the paper and will be presented elsewhere.) We assume that $M_{u^M}$ and $M_{d^M}$ are diagonal. The problem is now reduced to a diagonalization of a $4 \times 4$ matrix of the form
\begin{equation}
	\label{4x4u}
	\tilde{{\cal M}}_{u,k}=\left( \begin{array}{cc}
	M_u & M^{i4}_{q_L q^M_R} \\
	 M^{4j}_{u_R u^{M}_L}& m_{u^M,k}
	\end{array} \right) \,,
\end{equation}
where $i,j,k=1,2,3$ and where $m_{u^M,k}$ denotes the mass of th $k$th up mirror quark. Similarly, one has
\begin{equation}
	\label{4x4d}
	\tilde{{\cal M}}_{d,k}=\left( \begin{array}{cc}
	M_d & M^{i4}_{q_L q^M_R} \\
	 M^{4j}_{d_R d^{M}_L}& m_{d^M,k}
	\end{array} \right) \,.
\end{equation}

For simplicity, let us assume $m_{u^M,k}= m_{u^M}$ and $m_{d^M,k}=m_{d^M}$. In a recent work, Ref.~\cite{quarkmass} constructs phenomenogically the up and down-quark mass matrices $M_u$ and $M_d$ which can reproduce the known phenomenology of the CKM matrix and quark masses. These matrices turn out to be Hermitian and have real determinants. A simple calculation shows that the results are very similar to the one-generation case with similar quantities such as $r_u$ and $r_d$. Now, $m_u$ and $m_d$ appearing in Eqs.~(\ref{newru},\ref{newrd}) are truly the masses of the first generation quarks. Once again, we find $\theta_{Weak} \propto m_{\nu}/m_{u},m_{\nu}/m_{d}$.

%with the same conclusion as the one obtained for the one-generation case.

The EW $\nu_R$ model \cite{nur} was first conceived to provide a testable model of the seesaw mechanism by making right-handed neutrinos non-sterile and sufficiently "light" (i.e. with a mass $M_R$ proportional to the electroweak scale). These right-handed neutrinos do not come by themselves but are members of right-handed $SU(2)_W$ doublets which include right-handed mirror leptons. $SU(2)_W$ anomaly freedom dictates that one should also have doublets of right-handed mirror quarks. (The model includes per family $SU(2)_W$- singlets: $e_R$, $u_R$, $d_R$ for the SM fermions and $e^M_L$, $u^M_L$ and $d^M_L$ for the mirror fermions.) In fact, the SM and mirror sectors allow us to evade the Nielsen-Ninomiya no-go theorem \cite{nielsen} which forbids the chiral SM model to be put on the lattice.

It turns out that the ingredients contained in the EW $\nu_R$ model are precisely those that allow us to solve the strong CP problem. First by being "vector-like" (SM and mirror fermions), it allows us to set $\theta_{QCD}=0$ at tree level. Second, by mixing the left-handed SM lepton doublets with the right-handed mirror lepton doublets through the Higgs singlet fields, one obtains the neutrino Dirac mass $m_D$ which participates in the seesaw mechanism ($m_D^2/M_R$). This same mixing also operates in the quark sector giving rise to mixing between SM and mirror quarks in the mass matrices, which, in turn, contributes to the CP-violating parameter $Arg Det (M_u M_d)$ in an interesting way. It vanishes if $m_D$ goes to zero and is small ($< 10^{-10}$) because $m_D \ll M_R$ as in the seesaw mechanism. It is surprising that two seemingly unrelated phenomena find a common niche in the EW $\nu_R$ model.

I would like to thank Paul Frampton, Alfredo Aranda, T.c. Yuan and Goran Senjanovic for useful remarks.
%
%
%%%%%%%%%%%%%%%%%%%%
\appendix
%%%%%%%%%%%%%%%%%%%%
%
%
%%%%%%%%%%%%%%%%%%%%	
%%%%%%%%%%%%%%%%%%%%
%\FloatBarrier
%%%%%%%%%%%%%%%%%%%%
%
%
%
%%%%%%%%%%%%%%%%%%%%
%


\begin{thebibliography}{50}
\bibitem{theta}
R.~J.~Crewther, P.~Di Vecchia, G.~Veneziano and E.~Witten,
  %``Chiral Estimate of the Electric Dipole Moment of the Neutron in Quantum Chromodynamics,''
  Phys.\ Lett.\  {\bf 88B}, 123 (1979)
  Erratum: [{\it ibid.}   {\bf 91B}, 487 (1980)].
\bibitem{peccei}
R.~D.~Peccei and H.~R.~Quinn,
  %``CP Conservation in the Presence of Instantons,''
  Phys.\ Rev.\ Lett.\  {\bf 38}, 1440 (1977);
  S.~Weinberg,
  %``A New Light Boson?,''
  Phys.\ Rev.\ Lett.\  {\bf 40}, 223 (1978);
  F.~Wilczek,
  %``Problem of Strong p and t Invariance in the Presence of Instantons,''
  Phys.\ Rev.\ Lett.\  {\bf 40}, 279 (1978).
 \bibitem{mohapat}
H.~Georgi,
  %``A Model of Soft CP Violation,''
  Hadronic J.\  {\bf 1}, 155 (1978);
   R.~N.~Mohapatra and G.~Senjanovic,
  %``Natural Suppression of Strong p and t Noninvariance,''
  Phys.\ Lett.\  {\bf 79B}, 283 (1978);
  M.~A.~B.~Beg and H.-S.~Tsao,
  %``Strong P, T Noninvariances in a Superweak Theory,''
  Phys.\ Rev.\ Lett.\  {\bf 41}, 278 (1978); A more recent paper is
R.~N.~Mohapatra and A.~Rasin,
  %``A Supersymmetric solution to CP problems,''
  Phys.\ Rev.\ D {\bf 54}, 5835 (1996).

\bibitem{SBCP}
A.~E.~Nelson,
  %``Naturally Weak CP Violation,''
  Phys.\ Lett.\  {\bf 136B}, 387 (1984);
  S.~M.~Barr,
  %``Solving the Strong CP Problem Without the Peccei-Quinn Symmetry,''
  Phys.\ Rev.\ Lett.\  {\bf 53}, 329 (1984);
    S.~M.~Barr, D.~Chang and G.~Senjanovic,
  %``Strong CP problem and parity,''
  {\it ibid.}  {\bf 67}, 2765 (1991);
K.~S.~Babu and R.~N.~Mohapatra,
  %``A Solution to the Strong {CP} Problem Without an Axion,''
  Phys.\ Rev.\ D {\bf 41}, 1286 (1990);
   P.~H.~Frampton and D.~Ng,
  %``Strong and weak CP in a model with a new gauged U(1) symmetry,''
  {\it ibid.} {\bf 43}, 3034 (1991);
  P.~H.~Frampton and T.~W.~Kephart,
  %``Chiral Aspon model: An Alternative fully gauged model of CP violation,''
  {\it ibid.}  {\bf 47}, 3655 (1993);
   P.~H.~Frampton and T.~W.~Kephart,
  %``Method for constructing models with strong CP invariance,''
  Phys.\ Rev.\ Lett.\  {\bf 66}, 1666 (1991);
  P.~H.~Frampton, P.~Q.~Hung and M.~Sher,
  %``Quarks and leptons beyond the third generation,''
  Phys.\ Rept.\  {\bf 330}, 263 (2000).
%\bibitem{Nakamura:2010zzi} 
%\cite{Nakamura:2010zzi}
%  K.~Nakamura {\it et al.}  [Particle Data Group Collaboration],
  %``Review of particle physics,''
%  J.\ Phys.\ G {\bf 37}, 075021 (2010).
\bibitem{nur}
  P.~Q.~Hung,
  %``A Model of electroweak-scale right-handed neutrino mass,''
  Phys.\ Lett.\ B {\bf 649}, 275 (2007);
 % P.~Q.~Hung,
  %``Implications of right-handed neutrinos with electroweak-scale masses,''
  %Frascati Phys.\ Ser.\  {\bf 44}, 313 (2007)
  %arXiv:0706.2753 [hep-ph]];
   P.~Q.~Hung,
  %``Electroweak-scale mirror fermions, mu ---> e gamma and tau ---> mu gamma,''
  {\it ibid.}  {\bf 659}, 585 (2008);
  % P.~Q.~Hung,
  %``Consequences of Pati-Salam unification of electroweak-scale active nu(R) model: keV sterile neutrinos: four families,''
  %Nucl.\ Phys.\ B {\bf 805}, 326 (2008)
 % [arXiv:0805.3486 [hep-ph]];
   A.~Aranda, J.~Hernandez-Sanchez and P.~Q.~Hung,
  %``Implications of the discovery of a Higgs triplet on electroweak right-handed neutrinos,''
  JHEP {\bf 0811}, 092 (2008);
  V.~Hoang, P.~Q.~Hung and A.~S.~Kamat,
  %``Electroweak precision constraints on the electroweak-scale right-handed neutrino model,''
  Nucl.\ Phys.\ B {\bf 877}, 190 (2013).

  
  \bibitem{nielsen}
   H.~B.~Nielsen and M.~Ninomiya,
  %``Absence of Neutrinos on a Lattice. 1. Proof by Homotopy Theory,''
  Nucl.\ Phys.\ B {\bf 185}, 20 (1981)
  Erratum: [{\it ibid.}  {\bf 195}, 541 (1982)].
%\bibitem{hung2}
    
%\bibitem{cms}
\bibitem{125}
 V.~Hoang, P.~Q.~Hung and A.~S.~Kamat,
  %``Non-sterile electroweak-scale right-handed neutrinos and the dual nature of the 125-GeV scalar,''
  Nucl.\ Phys.\ B {\bf 896}, 611 (2015).

\bibitem{search}
  S.~Chakdar, K.~Ghosh, V.~Hoang, P.~Q.~Hung and S.~Nandi,
  %``Search for mirror quarks at the LHC,''
  Phys.\ Rev.\ D {\bf 93}, no. 3, 035007 (2016);
  S.~Chakdar, K.~Ghosh, V.~Hoang, P.~Q.~Hung and S.~Nandi,
  %``The search for electroweak-scale right-handed neutrinos and mirror charged leptons through like-sign dilepton signals,''
  {\it ibid.} {\bf 95}, no. 1, 015014 (2017).
\bibitem{muegamma}
P.~Q.~Hung, T.~Le, V.~Q.~Tran and T.~C.~Yuan,
  %``Lepton Flavor Violating Radiative Decays in EW-Scale $\nu_R$ Model: An Update,''
  JHEP {\bf 1512}, 169 (2015).
\bibitem{mu2e}
P.~Q.~Hung, T.~Le, V.~Q.~Tran and T.~C.~Yuan,
  %``Muon-to-Electron Conversion in Mirror Fermion Model with Electroweak Scale Non-Sterile Right-handed Neutrinos,''
  arXiv:1701.01761 [hep-ph].
\bibitem{quarkmass}
P.~Q.~Hung and T.~Le, in preparation.
%\bibitem{Ishimori:2010au} 
%\cite{Ishimori:2010au}
  %%CITATION = ARXIV:1003.3552;%%
  %297 citations counted in INSPIRE as of 10 Nov 2014

%\bibitem{Zhao:2012sqa} 
%\cite{Zhao:2012sqa}
\end{thebibliography}
\end{document}